# Universal Multichannel Coherence-Mediated Thermal Transport in Carbon Allotropes: Experimental Evidence and Scaling Laws

A. Jeżowski[1], A. I. Krivchikov[1,2], O. A. Korolyuk[2], V. V. Sagan[2], Y. V. Horbatenko[2], S.V. Cherednychenko[2], A.V. Dolbin[2], N.A. Vinnikov[2], O. Bezkrovnyi[1], D. Szewczyk[1*]

[1]Institute of Low Temperature and Structure Research, Polish Academy of Sciences, Wroclaw, Poland

[2]B. Verkin Institute for Low Temperature Physics and Engineering, National Academy of Sciences of Ukraine, Kharkiv, Ukraine

* corresponding author: d.szewczyk@intibs.pl

**Abstract**

Understanding thermal transport in structurally disordered carbon materials remains a fundamental challenge because conventional models based on a single coherence contribution cannot accurately describe heat conduction over a broad temperature range. In this work, the thermal conductivity of graphene oxide (GO), thermally reduced graphene oxide (trGO), GO–hydrated fullerene composites (GO–HyFn), expanded graphite (EG), and EG–multiwalled carbon nanotube (MWCNT) composites was measured between 4.5 and 300 K. Despite substantial differences in density, porosity, graphitic ordering, and microstructure, all materials exhibit glass-like temperature dependence of thermal conductivity.

The experimental results demonstrate that the conventional two-channel model, comprising a Peierls contribution and a single thermally activated coherence term, systematically fails to reproduce the observed temperature dependence. To resolve this discrepancy, we propose a generalized multichannel phenomenological model in which the coherence contribution consists of up to three Arrhenius-type channels with distinct activation energies. Depending on the structural state, two or three coherence channels are required to accurately describe the thermal conductivity.

A key finding is the reproducible hierarchy of activation energies across diverse carbon allotropes, following approximately ($E_2 \approx 4E_1$ and $E_3 \approx (12 \div 25)\ E_1$). While the channel amplitudes increase with material density, the activation energies remain nearly invariant, indicating universal vibrational energy scales. The proposed framework also consistently describes published data for other disordered carbon materials. These results provide experimental evidence for multichannel coherence-mediated heat transport and establish a unified phenomenological bridge between structural hierarchy with thermal transport in disordered carbon solids.

## 1. Introduction

Thermal conductivity is a fundamental property governing the performance of materials in applications ranging from microelectronics and thermoelectric energy conversion to thermal management and cryogenic technologies. Carbon-based materials are particularly attractive because their thermal conductivity can be tuned over several orders of magnitude through changes in crystal structure, dimensionality, density, porosity, and defect concentration. At low temperatures, where quantum statistics and intrinsic vibrational excitations dominate thermal transport, these materials provide an ideal platform for investigating the fundamental mechanisms of heat conduction [1-4].

Graphene and related carbon allotropes have attracted enormous attention because of their exceptionally high intrinsic in-plane thermal conductivity, among the highest reported for any material [4-7]. In practical systems, however, lattice defects, oxygen-containing functional groups, grain boundaries, stacking faults, and interlayer disorder strongly suppress heat transport. Consequently, thermal conductivity becomes highly sensitive to the degree of structural order, layer orientation, interflake coupling, and sample density, making graphene-derived materials particularly suitable for studying the relationship between atomic structure and thermal transport [8-10].

The microscopic origin of heat transport in structurally disordered solids has recently become one of the central topics in condensed-matter physics. Within the traditional picture, vibrational excitations are classified into propagons, diffusons, and locons, depending on their degree of spatial localization and ability to transport thermal energy. While this classification has been widely adopted for amorphous materials, the relative contribution of these vibrational excitations to thermal transport in amorphous carbon and related disordered carbon allotropes remains actively debated [11-12]. Recent theoretical studies further demonstrate that quantum statistics profoundly influence thermal conduction at low temperatures, modifying the relative importance of different vibrational channels and establishing new transport regimes in low-dimensional amorphous solids [12].

Historically, thermal transport in amorphous materials has been described by the harmonic theory of Allen and Feldman, in which heat is carried by harmonic vibrational modes undergoing random energy diffusion rather than by propagating phonons [13]. More recently, this description has been extended by unified theories of thermal transport that treat heat conduction as the combined contribution of propagating phonons and coherence-mediated transport between vibrational eigenstates. In this multichannel framework, particle-like and wave-like transport coexist and continuously evolve with increasing structural disorder, providing a unified description of thermal conductivity in both crystalline and amorphous solids [14-15].

Understanding how atomistic disorder governs thermal conductivity remains one of the major unresolved problems in materials physics. Although numerous experimental studies have demonstrated that increasing structural disorder generally suppresses thermal transport, establishing a quantitative relationship between microscopic structure and macroscopic thermal conductivity has proven challenging. Recent work by Iwanowski, Csányi, and Simoncelli [16] has shown that disorder-induced variations of thermal conductivity in network solids can be quantitatively predicted using the bond-network entropy, an atomistic descriptor characterizing the heterogeneity of the atomic-bond topology. This descriptor correlates directly with the smoothness of the vibrational density of states and provides a quantitative measure of the structural length scales controlling heat transport in disordered materials. These advances establish a direct connection between local atomic topology, vibrational spectra, and thermal conductivity.

Reduced graphene oxide (rGO) and thermally reduced graphene oxide (trGO) constitute intermediate structures between graphene oxide and crystalline graphite. Thermal reduction removes oxygen-containing functional groups and partially restores the $sp^2$ carbon network, thereby increasing thermal conductivity while simultaneously introducing vacancies, lattice distortions, and stacking disorder [17-22]. Likewise, expanded graphite (EG), multiwalled carbon nanotubes (MWCNTs), hydrated fullerene (HyFn) composites, and nanodiamond ceramics exhibit markedly different dimensionalities, defect landscapes, and interfacial structures, making them an ideal family of materials for investigating how structural disorder modifies competing heat-transport channels.

In this work, we investigate the low-temperature thermal conductivity of several nanostructured carbon allotropes, including graphene oxide (GO), GO–hydrated fullerene composites (GO–HyFn), expanded graphite (EG)–multiwalled carbon nanotube (MWCNT) composites. Steady-state thermal conductivity measurements performed between 4 and 300 K are used to examine the influence of density, compaction pressure, annealing, and nanostructure on heat transport. The experimental results are analyzed within the framework of contemporary multichannel thermal transport, demonstrating how structural disorder controls the balance between propagating and non-propagating vibrational heat carriers in carbon-based materials.

## 2. Materials and Experimental Methods

### 2.1 Thermal conductivity measurements

The thermal conductivity was measured by a steady-state uniaxial heat-flow technique in the temperature range 4.5–100 K using a specially designed cryogenic apparatus [23]. Measurements were performed under high-vacuum conditions, with the specimen thermally isolated by multiple radiation shields. The temperature gradient along the sample was monitored using calibrated constantan–manganin differential thermocouples positioned several millimeters apart, while a miniature electrical heater attached to one end of the specimen generated a stationary heat flux. To minimize parasitic heat losses through electrical leads, the radiation shield closest to the sample was maintained at a stabilized temperature. The overall uncertainty of the thermal conductivity measurements did not exceed 15%, including approximately 5% associated with the uncertainty in thermocouple spacing, whereas the reproducibility of the experimental data was better than ±2%. A detailed description of the experimental setup and measurement procedure is given in Ref. [23].

### 2.2 Materials

Two families of carbon-based materials exhibiting different degrees of structural disorder and dimensionality were investigated: (i) graphene oxide (GO)-based materials, including thermally reduced graphene oxide (trGO) and GO–hydrated fullerene (GO–HyFn) composites, and (ii) expanded graphite (EG) and EG–multiwall carbon nanotube (MWCNT) composites. These systems provide a broad range of microstructures, densities, and defect concentrations, making them suitable model materials for studying multichannel thermal transport.

**Graphene oxide and thermally reduced graphene oxide**

Graphene oxide was synthesized from high-purity graphite using a modified Hummers method at the B. Verkin Institute for Low Temperature Physics and Engineering, National Academy of Sciences of Ukraine, (Kharkiv, Ukraine) [24]. Oxidation introduced hydroxyl, epoxy, and carboxyl functional groups, disrupted the extended $sp^2$ carbon network, and increased the interlayer spacing between graphene sheets [25].

Thermal exfoliation was performed under vacuum ($\sim 10^{-3}$ mmHg) by heating the oxidized graphite to 300 °C, producing samples denoted GO_300_S3 and S5). Additional thermal reduction at 300, 500, and 700 °C yielded a series of thermally reduced graphene oxide (trGO) samples with progressively decreasing oxygen content and increasing structural ordering. Simultaneously, powder compaction under loads of 0.75, 1.0 and 1.25 GPa allowed systematic variation of the sample density and interlayer contact between graphene sheets. Consequently, both thermal reduction and mechanical densification independently modified the defect landscape governing heat transport.

Energy-dispersive X-ray spectroscopy (EDS) showed that the trGO_300 samples contained approximately 83.6 at.% carbon and 14.0 at.% oxygen, with less than 2.4 at.% residual impurities [26] . Increasing the reduction temperature progressively removed oxygen-containing functional groups, reduced the concentration of oxygen-related defects, and partially restored the graphitic $sp^2$ framework [25-26].

**GO–hydrated fullerene composites**

Hydrated fullerene (HyFn) was synthesized by cryogenic co-condensation of $C_{60}$ and water vapor onto a substrate maintained at 90 K. After melting the condensate, a transparent aqueous dispersion containing hydrated fullerene complexes with characteristic diameters of approximately 1.6–1.8 nm was obtained.

GO–HyFn composites were prepared by dispersing GO in water followed by the addition of an aqueous HyFn solution. Hydrated fullerene species adsorbed onto the graphene oxide surface through π–π interactions and hydrophobic forces. 0.5 mL of HyFn solution (10 mg $L^{-1}$) was mixed with 1 g of trGO powder, corresponding to a HyFn concentration of approximately $5 \times 10^{-4}$ wt.%. The suspension was magnetically stirred for 4 h at room temperature to promote penetration of HyFn into pores, folds, and interlayer galleries before drying under ambient conditions.

For thermal conductivity measurements, powders were compacted into pellets under loads of 1.0 and 1.25 GPa, providing samples denoted GO-HyFn_S# (# = 1–2) with controlled densities and interfacial contact areas for thermal conductivity measurements.

*Structural characterization*

The morphology of the GO-based materials was investigated by transmission electron microscopy (TEM) using a Philips CM-20 SuperTwin microscope operated at an accelerating voltage of 160 kV.

Figure 1 shows representative TEM images of the GO–HyFn_S2 composite. Low-magnification images (Fig. 1a) reveal a loose network of thin, partially transparent graphene oxide sheets without clearly distinguishable fullerene particles. The selected-area electron diffraction (SAED) pattern (Fig. 1b) exhibits weak diffraction rings that can be assigned to the fullerene phase. Higher-magnification images (Fig. 1c, d) demonstrate wrinkled and partially restacked GO lamellae containing only a few small regions of enhanced contrast that are consistent with isolated HyFn clusters or very small aggregates. No large fullerene-rich domains are observed, indicating a nearly homogeneous distribution of hydrated fullerene throughout the graphene oxide framework.

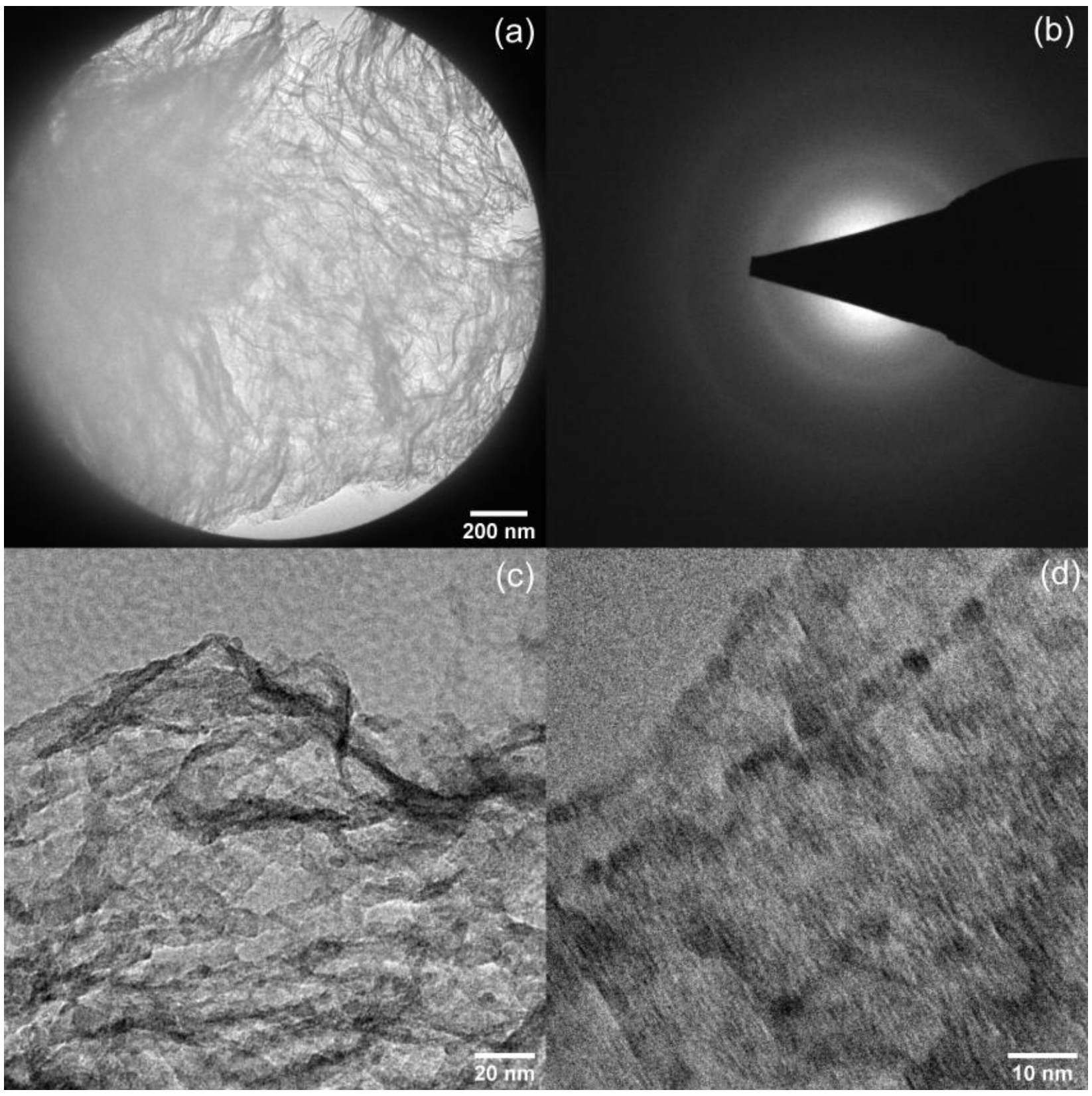

**Fig. 1.** Low-magnification TEM image of the GO–HyFn_S2 powder sample (a). Selected-area electron diffraction (SAED) pattern (b), and representative medium- and high-magnification TEM images of GO–HyFn_S2 (c, d).

## Expanded graphite and EG–MWCNT composites

Expanded graphite and EG–MWCNT composites were synthesized at the Chuiko Institute of Surface Chemistry, National Academy of Sciences of Ukraine (Kiev, Ukraine) [27]. Expanded graphite was produced by thermal-shock exfoliation of graphite intercalation compounds, yielding highly porous networks containing slit-shaped and cylindrical pores with characteristic dimensions of 2–40 nm. Raman spectroscopy, positron annihilation spectroscopy, and low-temperature gas adsorption confirmed the hierarchical porous structure and pronounced structural heterogeneity of the material.

Multiwalled carbon nanotubes (MWCNTs) were synthesized by catalytic chemical vapor deposition and exhibited diameters of 10–20 nm, specific surface areas of 200–400 $m^2\ g^{-1}$, and bulk densities of 20–40 g $dm^{-3}$. Expanded graphite samples (#6-EleChem$^{II–III}$, #7-Chem$^{I}$, and #8-Chem$^{I}$) together with EG–MWCNT composites (#13-Chem$^{I}$-3%CNTs and #14-Chem$^{I}$-3%CNTs) were prepared using different oxidation routes and graphite intercalation stages. The designations *Chem* and *EleChem* correspond to chemical and electrochemical oxidation, respectively, whereas the indices I and II–III indicate first-stage and mixed second-/third-stage graphite intercalation compounds. The CNT-modified samples contained 3 wt.% MWCNTs.

Different oxidation routes and CNT incorporation resulted in distinct structural states of expanded graphite, which, in turn, affected the thermal transport properties of the composites.

The main characteristics of the investigated samples are summarized in Table 1.

**Table 1.** The base information about the sample for thermal conductivity experiments.

| Sample | Material | Compacted by pressure (GPa) | Annealing temperature (°C) | Samples Sizes (mm×mm×mm) |
|---|---|---|---|---|
| Compacted GO | | | | |
| Series I | | | | |
| GO-HyFn_S1 | GO -HyFn | 1.0 | 300 | 10×6×2 |
| GO-HyFn_S2 | GO -HyFn | 1.25 | 300 | 10×6×2 |
| GO_300ºC_S3 | GO | 1.25 | 300 | 10×6×2 |
| Series II | | | | |
| GO_300°C_6t_S5 | GO | 0.75 | 300 | 10×6×2 |
| GO_300°C_8t_S5 | GO | 1.0 | 300 | 10×6×2 |
| GO_300°C_10t_S5 | GO | 1.25 | 300 | 10×6×2 |
| GO_500°C_10t | GO | 1 | 500 | 6×4×2 |
| GO_700°C_10t | GO | 1 | 700 | 6×4×2 |
| EG and EG-MWCNT composites | | | | |
| #6-Elechem[II-III] $\rho$ = 0.357 g/cm$^3$ | EG | - | - | 15×6×0.50 |
| #7-Chem[I] $\rho$ = 0.73 g/cm$^3$ | EG | - | - | 15×6×0.65 |
| #8-Chem[I] $\rho$ = 1.3 g/cm$^3$ | EG | - | - | 15×6×0.67 |
| #13-Chem[I]-3%CNTs $\rho$ = 0.966 g/cm$^3$ | EG-3%CNTs | - | - | 15×6×0.50 |
| #14-Chem[I]-3%CNTs $\rho$ = 1.62 g/cm$^3$ | EG-3%CNTs | - | - | 15×6×0.42 |

The combined variation of oxidation route, thermal reduction, densification, and nanotube incorporation produced carbon materials spanning a wide range of structural order, defect concentration, pore morphology, and interfacial density. This diversity provides an excellent experimental platform for investigating how competing propagating and non-propagating vibrational channels govern thermal transport in carbon allotropes.

### 2.3 Structural hierarchy of the investigated carbon allotropes

The investigated materials were deliberately selected to represent a broad hierarchy of carbon structures spanning different dimensionalities, degrees of crystallinity, defect concentrations, pore architectures, and interfacial connectivity. Such structural diversity provides an experimental platform for examining how different vibrational heat-transport channels evolve with increasing structural disorder.

Graphene oxide (GO) represents the most chemically disordered system in the present study. Oxidation introduces hydroxyl, epoxy, and carboxyl functional groups, fragments the extended sp$^2$ network, increases the interlayer spacing, and creates a high density of structural defects that strongly modify the vibrational spectrum. Progressive thermal reduction partially restores the graphitic framework while simultaneously generating vacancies and lattice distortions. Consequently, thermally reduced graphene oxide (trGO) samples provide a controlled series with systematically varying structural order.

The incorporation of hydrated fullerene (HyFn) into the graphene oxide framework introduces an additional level of structural complexity. Fullerene molecules occupy pores and interlayer regions, creating new heterogeneous interfaces without significantly altering the overall morphology of the graphene sheets. These interfaces modify the coupling between neighboring graphene layers and introduce additional pathways for vibrational energy transfer.

Expanded graphite (EG) occupies an intermediate structural regime between layered graphite and highly porous carbon networks. Thermal exfoliation produces a hierarchical pore structure composed of interconnected graphene lamellae with varying orientation and contact quality. The incorporation of multiwalled carbon nanotubes (MWCNTs) further modifies the interfacial topology by forming conductive bridges between adjacent graphite flakes, thereby altering both the connectivity of the carbon framework and the efficiency of vibrational energy transfer across interfaces.

Overall, the investigated materials span the transition from highly defective and weakly connected graphene oxide to densely compacted graphitic ceramics through intermediate states characterized by controlled defect density, porosity, and interface density. Such a systematic structural hierarchy makes it possible to examine how propagating phonons and non-propagating vibrational excitations evolve across different carbon allotropes and to identify the structural parameters governing multichannel thermal transport.

From the perspective of the modern multichannel theory of heat transport, increasing structural disorder is expected not only to enhance phonon scattering but also to progressively modify the relative contributions of propagating phonons, diffuson-like vibrational modes, and coherence-mediated transport. The investigations enable an experimental assessment of how changes in atomic-scale disorder and interfacial topology redistribute the contributions of different heat-transport channels in carbon allotropes.

### 3. Multichannel model of thermal transport

Heat transport in structurally complex solids cannot, in general, be described solely within the conventional phonon-gas model. Structural disorder, interfaces, and reduced dimensionality progressively suppress long-lived propagating phonons and promote additional mechanisms of vibrational energy transfer. Consequently, thermal conductivity arises from multiple transport channels whose relative contributions depend on temperature and the degree of structural order.

Historically, non-propagating thermal transport in disordered solids was first described within the harmonic theory of Allen and Feldman [13, 28-30], which classifies vibrational excitations into propagons, diffusons, and locons. In this framework, low frequency propagons transport heat similarly to phonons in crystals, diffusons carry energy through harmonic coupling between spatially extended vibrational modes, and highly localized locons contribute negligibly to thermal conductivity.

A complementary description was proposed by Cahill and Pohl [31], who introduced the minimum thermal conductivity model for highly disordered solids. This approach assumes that heat is transferred by diffusively propagating vibrational excitations with mean free paths approaching interatomic distances, thereby providing an estimate of the lower limit of thermal conductivity above the plateau region.

These concepts have been unified within the quantum-mechanical theory of thermal transport based on the Wigner transport equation developed by Simoncelli et al., and others [14-15, 32-34], particle-like phonon propagation and wave-like coherence between vibrational eigenstates. The theory therefore provides a description of heat transport in crystalline, defective, and amorphous solids without introducing separate physical models for different classes of materials.

Within this framework, thermal conductivity is naturally decomposed into two physically distinct contributions,

$$\kappa(T) = \kappa_P(T) + \kappa_C(T), \qquad (1)$$

where $\kappa_P(T)$ is the Peierls contribution arising from the propagation of phonon wave packets, while $\kappa_C(T)$ represents the coherence contribution originating from quantum coupling between vibrational states.

The Peierls contribution dominates when long-lived propagating phonons exist, particularly at low temperatures where phonon populations are governed by Bose–Einstein statistics and anharmonic scattering remains weak. Increasing structural disorder, however, progressively reduces phonon lifetimes and shortens their mean free paths, thereby decreasing the relative importance of this transport channel.

The coherence contribution becomes increasingly significant when disorder, defects, or complex atomic topology produce strong spectral overlaps between vibrational modes. Rather than representing ballistic phonon transport, this channel corresponds to wave-like energy transfer mediated by coherent coupling between neighboring vibrational eigenstates. In highly disordered solids, this mechanism continuously evolves toward the diffuson transport originally described by Allen and Feldman, thereby establishing a direct connection between the modern multichannel theory and the classical description of amorphous thermal transport [13, 28-37].

Although the Wigner formalism enables first-principles calculations of both transport channels, quantitative evaluation of the coherence contribution requires computationally demanding lattice-dynamical calculations and knowledge of the complete vibrational spectrum [32]. For experimental analysis, it is therefore advantageous to employ a phenomenological representation that captures the observed temperature dependence while retaining the physical interpretation of the coherence channel.

Previous investigations of structurally disordered materials, including triphenyl phosphite [38], clathrate compounds [39], amorphous polymers [40], ABS/trGO composites [41], orientationally disordered molecular crystals [42], and solid solutions of complex crystals [43] have demonstrated that the coherence contribution is accurately described over a broad temperature range by an Arrhenius-type expression,

$$\kappa_C(T) = \kappa_0 \exp(-E/T), \tag{2}$$

where $\kappa_0$ is the limiting magnitude of the coherence contribution at elevated temperatures and $E$ is an effective activation energy characterizing the dominant non-propagating vibrational excitations, expressed in Kelvin.

The present work adopts this phenomenological multichannel framework to analyze thermal transport in carbon allotropes with different degrees of structural order. Rather than assigning individual vibrational modes to specific categories, the model separates the experimentally measured thermal conductivity into propagating and coherence-mediated contributions, allowing the evolution of competing heat-transport channels to be quantified as a function of density, defect concentration, porosity, and microstructural organization.

## 4. Results and discussion

### 4.1. Low-temperature thermal conductivity of graphene oxide

Figure 2 presents the temperature dependence of the thermal conductivity, $\kappa(T)$, of the graphene oxide (GO) samples measured between 4.5 and 300 K. The investigated materials include the as-prepared GO sample (GO_300ºC_S3, Series I) together with a second series of pellets (GO_300ºC_6t_S5, GO_300ºC_8t_S5, and GO_300ºC_10t_S5) obtained by successive compaction of the same powder under loads of 0.75, 1.0 and 1.25 GPa without intermediate thermal treatment. This series provides a direct means of evaluating the influence of mechanical densification on heat transport while minimizing variations in chemical composition and defect concentration.

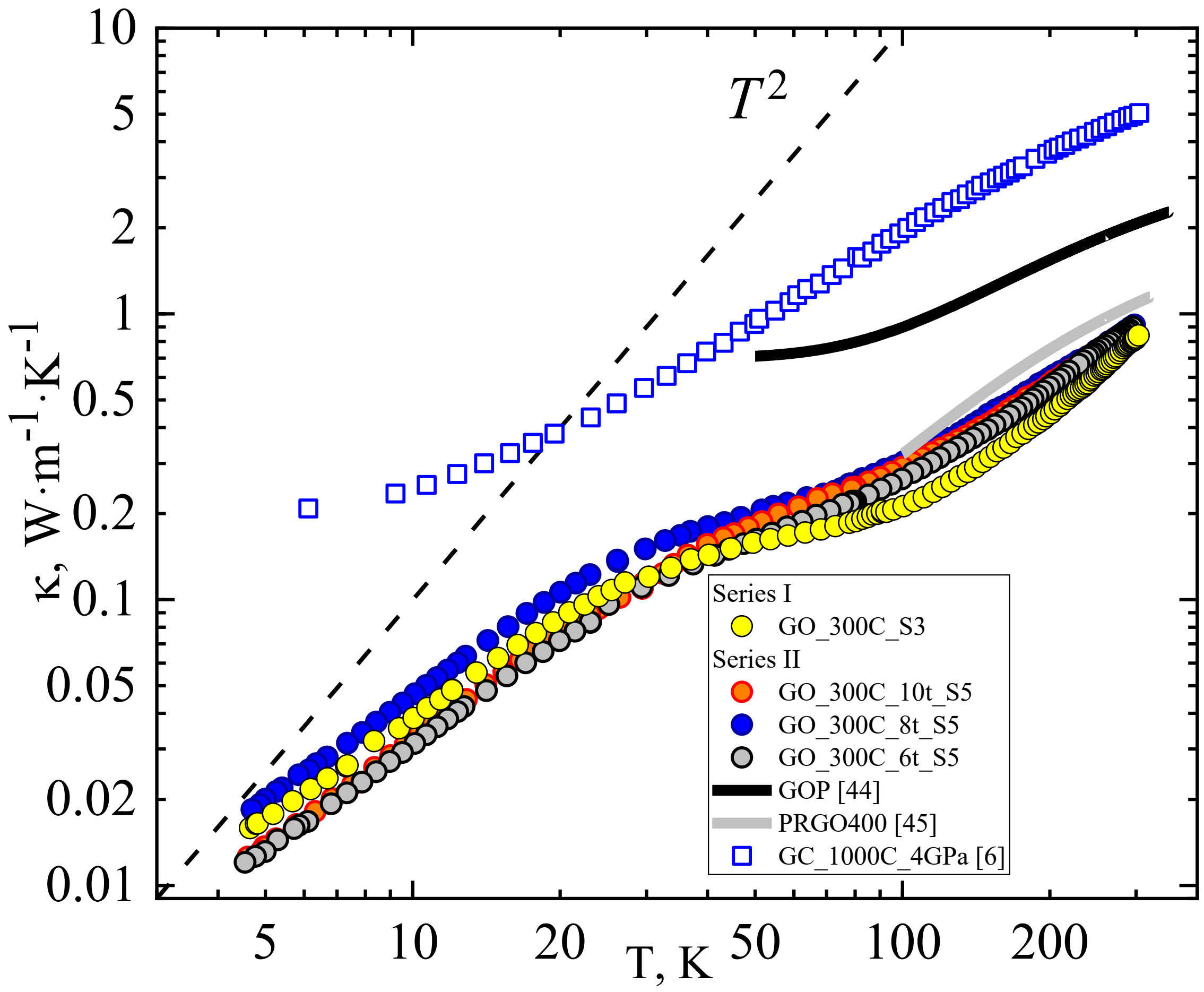

**Fig. 2.** The temperature dependence of the thermal conductivity of graphene oxide samples, symbols: series I: (●) - GO_300ºC_S3; series II: (•) - GO_300ºC_6t_S5, (●) - GO_300ºC_8t_S5, (●) - GO_300ºC_10t_S5 and literature data [4] for graphene ceramic: (□) – GC_1000°C_4GPa. Solid lines represent literature data for graphene oxide paper GOP [44] and partly-reduced graphene oxide films (PRGO400) [45]. The dashed straight line indicates the dependence $\kappa(T) \sim T^2$.

All investigated samples exhibit low thermal conductivity throughout the entire temperature range together with a monotonic increase of $\kappa(T)$ upon heating. No maximum associated with crystalline phonon transport is observed. Instead, the thermal conductivity displays a smooth temperature dependence typical of structurally disordered materials. The typical low-temperature plateau commonly observed in glasses is not reached within the present measurements and is therefore expected to lie below the lowest experimental temperature of 4.5 K.

Mechanical compaction has only a moderate influence on thermal conductivity at low temperatures. Below approximately 40 K, the $\kappa(T)$ curves do not have large difference, indicating that densification only weakly affects heat transport in this temperature region. At higher temperatures, however, the influence of compaction becomes progressively more pronounced. The sample compacted under a load of 10 tons exhibits the highest thermal conductivity, exceeding that of the initial GO_300ºC_S3 sample by approximately 37% near 100 K. All samples further exhibit a broad region of weak temperature dependence extending approximately from 40 to 90 K. With increasing compaction load, this broad plateau shifts slightly toward lower temperatures, suggesting that densification modifies the characteristic temperature scale governing heat transport.

For comparison, Fig. 2 also includes published data for graphene oxide paper (GOP) [44] and partially reduced graphene oxide (PRGO400) films [45] and literature data [6] for graphene ceramic GC. Although the absolute values of the thermal conductivity differ somewhat, the present GO samples exhibit temperature dependences closely resembling those reported for other graphene-based materials. The measured thermal conductivity is comparable to that of PRGO400 above 100K and lower than that of graphene oxide paper over the corresponding temperature intervals. The temperature dependence of the graphene-like ceramics fabricated by the high isostatic pressure method at a temperature of 1000°C and at pressure 4GPa is like the dependence of other graphene oxide materials.

The high-pressure graphene ceramic represents the densest material investigated in this work. Hot isostatic pressing substantially reduces porosity, increases the contact area between graphene layers, and enhances structural continuity while preserving the layered carbon framework. These samples therefore provide a useful reference for evaluating the influence of compaction and interfacial coupling on thermal transport.

The dashed line in Fig. 2 represents the quadratic dependence, $\kappa(T) \sim T^2$, expected for conventional low-temperature phonon transport in amorphous solids. It is evident that none of the investigated graphene oxide samples follows this behavior at the lowest temperatures Instead, the measured thermal conductivity increases considerably slower than $T^2$, indicating that the thermal transport cannot be described by the conventional low-temperature approximation over the investigated temperature range. This deviation constitutes the first indication that heat transport in graphene oxide involves more complex vibrational processes than those captured by a single propagating-phonon contribution and motivates the multichannel analysis developed in the following sections.

### 4.2. Breakdown of the conventional two-channel description

To determine whether the generalized two-channel model introduced in Section 3 adequately describes the thermal conductivity of graphene-based materials, the experimental data were first analyzed using Eq. (1), in which the thermal conductivity is represented as the sum of a Peierls contribution, $\kappa_P(T)$, and a single thermally activated coherence contribution, $\kappa_C(T)$, given by Eq. (2). Figure 3 presents this analysis for the compacted graphene oxide sample GO_300°C_6t_S5 together with GO_500°C_10t and GO_700°C_10t.

At low temperatures, the Peierls contribution contains both the temperature-dependent component associated with resonant scattering by two-level-system (TLS) excitations and quasi-local vibrations, together with the temperature-independent plateau thermal conductivity, $\kappa_{pl}$ [40]. Because the present measurements extend only down to 4.5 K, the parameters describing the TLS contribution cannot be determined reliably. Consequently, the Peierls term was approximated by the experimentally observed plateau conductivity, $\kappa_P = \kappa_{pl}$, where the corresponding values of $\kappa_{pl}$ are indicated by the horizontal dotted lines in Fig. 3.

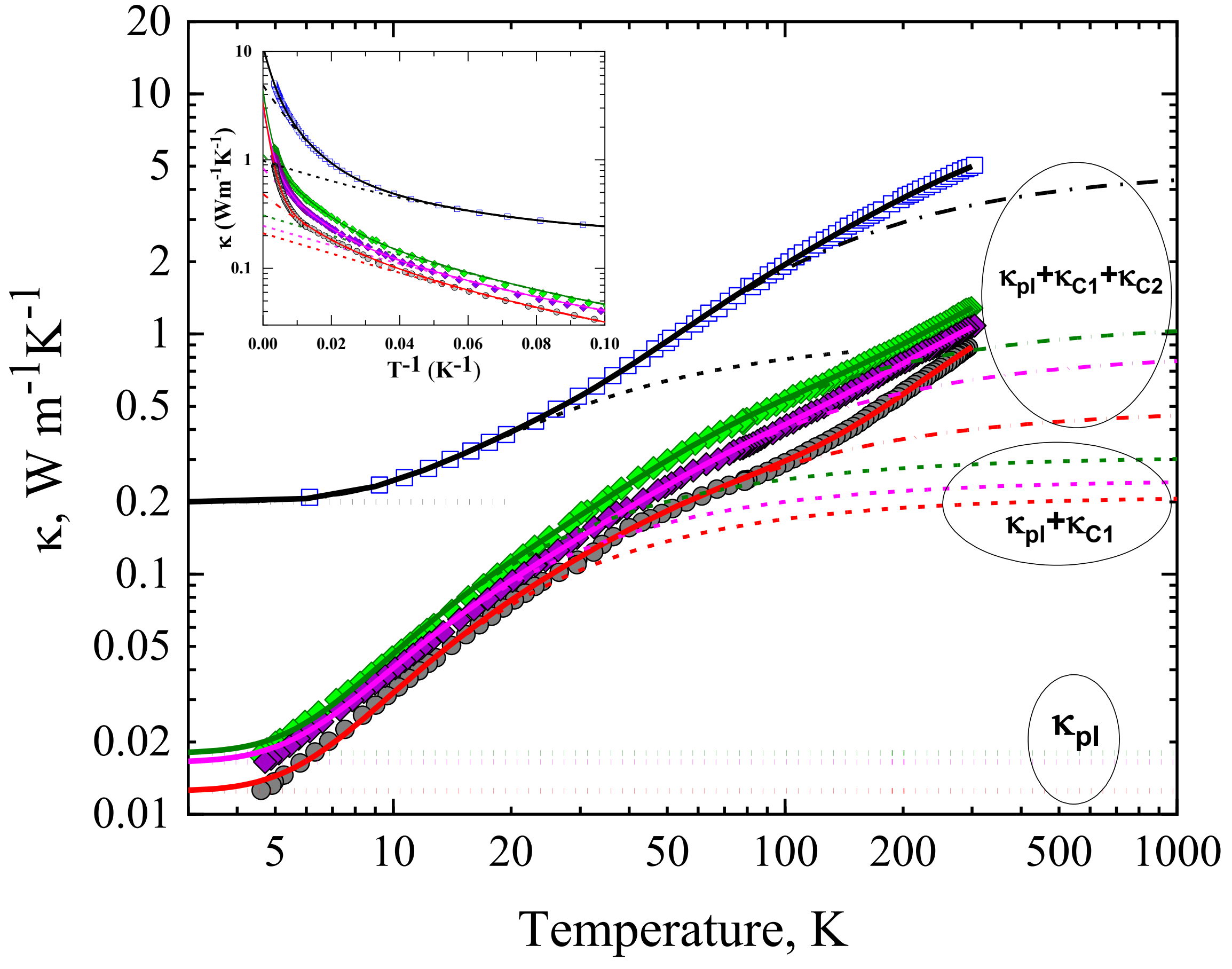

**Fig.3.** The temperature dependence of the thermal conductivity of graphene oxide sample (●) -GO_300ºC_6t_S5, series I, and graphene ceramics (♦) - GO_500°C_10t, (♦) - GO_700°C_10t and literature data [4] for graphene ceramic: (□) – GC_1000°C_4GPa. Lines are fitting curves: dot-lines are $\kappa_{pl}$, dash-lines are $\kappa_{pl} + \kappa_{C1}(T)$, dash-dot-lines are $\kappa_{pl} + \kappa_{C1}(T) + \kappa_{C2}(T)$, and solid lines are $\kappa_{pl} + \kappa_{C1}(T) + \kappa_{C2}(T) + \kappa_{C3}(T)$. The colors of the lines correspond to those of the symbols. The inset shows the experimental data together with the fitted curves plotted in semilogarithmic coordinates ($\kappa$ - $\kappa_{pl}$) *vs* $1/T$.

Under this approximation, the conventional multichannel model predicts that the temperature dependence above the plateau should be governed by a single activated coherence contribution described by Eq. (2). As shown by the dashed curves in Fig. 3, this description reproduces the experimental thermal conductivity only within the low-temperature region below approximately 25 K. Above this temperature, systematic deviations become evident for all investigated samples. Rather than approaching the measured thermal conductivity, the calculated curves increasingly underestimate the experimental values with increasing temperature.

The failure of the single activated contribution is illustrated more clearly in the inset of Fig. 3, where the quantity ($\kappa$ - $\kappa_{\text{pl}}$) is plotted as a function of ($1/T$) in semilogarithmic coordinates. Following the analysis developed in Refs. [38-43], the low-temperature data exhibit the expected linear behavior, allowing the first activated contribution $\kappa_{C1}(T)$ to be characterized by the pre-exponential factor $\kappa_{01}$ and the characteristic energy $E_1$, obtained from the intercept and slope of the linear dependence, respectively. However, above approximately 25 K, the experimental data deviate systematically from this linear behavior, demonstrating that a single Arrhenius contribution is insufficient to describe the thermal conductivity over the full experimental temperature range.

This systematic deviation constitutes the principal experimental result of the present work. It demonstrates that the coherence contribution predicted by the conventional two-channel model cannot be represented by a single characteristic energy. Instead, experimental thermal conductivity requires the existence of additional thermally activated coherence channels that become progressively important as the temperature increases.

To account for this behavior, the coherence contribution is represented as the sum of several activated components with different characteristic energies. The thermal conductivity can therefore be expressed as:

$$\kappa(T) = \kappa_{\text{pl}} + \kappa_{C1}(T) + \kappa_{C2}(T) + \kappa_{C3}(T) =$$

$$\kappa_{\text{pl}} + \kappa_{01}\exp(-E_1/T) + \kappa_{02}\exp(-E_2/T) + \kappa_{03}\exp(-E_3/T), \quad (3)$$

where $\kappa_{0i}$ is the pre-exponential factor, $E_i$ is the characteristic activation energy of the $i$-the coherence channel. Here, $i$ varies from 1 to 3 and corresponds to Arrhenius-type heat-transfer contributions arising from different excitation mechanisms.

For the graphene oxide investigated here and graphene ceramic samples [4], three coherence channels are required to reproduce the measured thermal conductivity over the entire temperature interval from 4.5 to 300 K. The corresponding fitting curves are shown in Fig. 3: the dash-dotted lines include the first two coherence channels, while the solid curves incorporate all three contributions. The resulting agreement with the experimental data is excellent throughout the investigated temperature range.

The extracted activation energies exhibit a remarkable degree of universality. The first characteristic energy, $E_1 \approx 23$ K, is nearly identical for all investigated samples. The second characteristic energy is approximately four times larger ($E_2 \approx 4E_1$), whereas the third is approximately twenty-four times greater ($E_3 \approx 24E_1$). The corresponding pre-exponential factors and activation energies are summarized in Table 2.

The necessity of introducing three thermally activated coherence channels demonstrates that heat transport in structurally complex graphene-based materials is considerably richer than predicted by the conventional two-channel description. Rather than representing a single activated process, the coherence contribution consists of several distinct energy scales that together govern thermal transport over the broad temperature interval extending from the low-temperature plateau to room temperature. The generality of this multichannel description is examined in the following subsection using a wider range of structurally diverse carbon allotropes.

### 4.3. Experimental evidence for multiple coherence channels

The analysis presented in the previous section demonstrates that the conventional two-channel description, consisting of a Peierls contribution and a single activated coherence term, is insufficient to reproduce the thermal conductivity of graphene oxide over the entire investigated temperature range. An important question therefore arises: Is this behavior specific to a particular sample, or does it represent a more general feature of structurally disordered carbon materials?

To address this question, the proposed multichannel model, Eq. (3), was applied to a broader family of carbon allotropes exhibiting markedly different microstructures and thermal conductivities. Figure 4 compares the temperature dependences of the thermal conductivity for graphene oxide (GO_300ºC_S3 and GO_300ºC_10t_S5) (symbols as in Fig. 2), hydrated-fullerene composites (GO–HyFn_S1 and GO–HyFn_S2), amorphous $SiO_2$ [31,46], and reticulated glassy carbon (rGC) [47], together with the corresponding fitting curves calculated using Eq. (3).

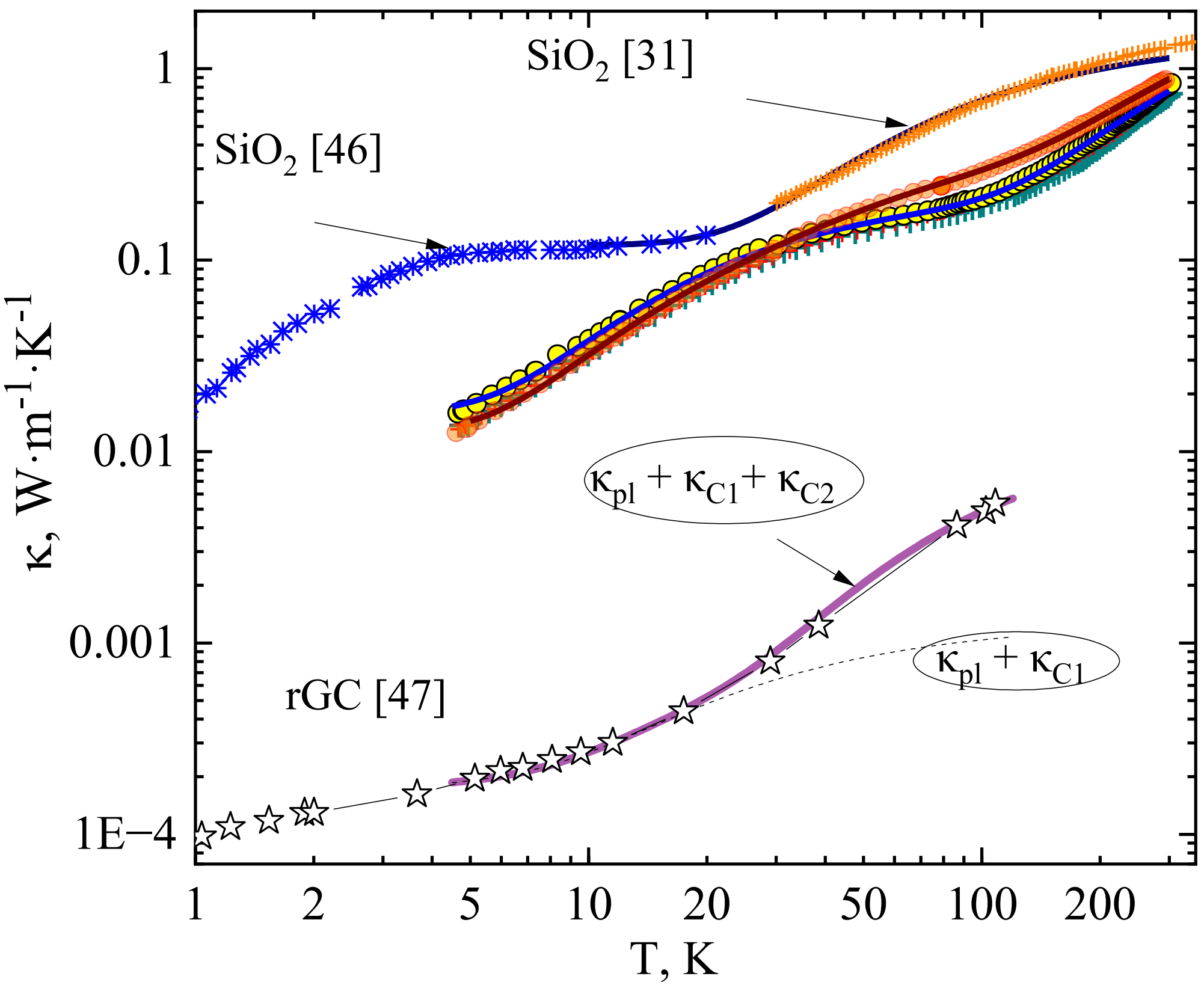

**Fig. 4.** Temperature dependences of the thermal conductivity of GO_300ºC_S3 (●), GO_300ºC_10t_S5 (●), GO–HyFn_S1 (×), and GO–HyFn_S2 (×). For comparison, data for amorphous silica $SiO_2$ [31, 46] (✳) and rGC (★) [47] are also shown. The solid colored lines

represent fits to the experimental data using Eq. (3), while the dashed line corresponds to $\kappa(T) = \kappa_{\text{pl}} + \kappa_{C1}(T)$ for rGC. The fitting parameters are listed in Table 2.

The GO-based materials exhibit remarkably similar thermal conductivities despite differences in compaction and chemical modification. In particular, the two hydrated-fullerene composites display thermal conductivities that are nearly identical to those of the parent graphene oxide samples throughout the investigated temperature range. Within the experimental uncertainty, the incorporation of hydrated fullerene therefore produces no measurable change in the thermal transport properties, indicating that the extremely small HyFn concentration is insufficient to modify the dominant vibrational transport processes.

The comparison with amorphous $SiO_2$ reveals a fundamentally different type of thermal-conductivity evolution. Silica exhibits classical glass-like behavior, characterized by an approximately quadratic increase of $\kappa(T)$ at low temperatures, a well-defined low-temperature plateau, between approximately 6 and 11 K, and a gradual increase toward saturation at higher temperatures. In contrast, the graphene-based materials show a considerably weaker temperature dependence at the lowest temperatures together with only a broad inflection extending over several tens of Kelvin at elevated temperatures, rather than a pronounced plateau. These observations immediately suggest that the underlying mechanisms governing heat transport differ substantially from those operating in conventional oxide glasses.

This distinction is further confirmed by the fitting analysis. Whereas the thermal conductivity of amorphous $SiO_2$ is accurately reproduced using a single activated coherence contribution ($\kappa(T) = \kappa_{\text{pl}} + \kappa_{C1}(T)$), none of the investigated graphene-based materials can be described satisfactorily within this approximation. Instead, excellent agreement with the experimental data is achieved only when the coherence contribution is represented by several activated channels according to Eq. (3).

Reticulated glassy carbon provides particularly compelling evidence for the generality of this behavior. Although its thermal conductivity is nearly two orders of magnitude lower than that of graphene oxide, its temperature dependence closely resembles that of the GO-based materials. As in graphene oxide, a single activated coherence contribution reproduces the experimental data only at low temperatures, whereas the complete temperature dependence requires additional activated channels. Thus, the necessity of multiple coherence contributions appears to be an intrinsic feature of structurally heterogeneous carbon networks.

The fitting parameters summarized in Table 2 reveal several remarkable regularities. The plateau conductivity, $\kappa_{\text{pl}}$, differs by almost three orders of magnitude among the investigated materials, ranging from amorphous $SiO_2$ to reticulated glassy carbon. Likewise, the amplitude of the first coherence channel, $\kappa_{01}$, varies by more than three orders of magnitude. In contrast, the characteristic activation energies $E_i$ remain remarkably stable. For all graphene oxide samples and reticulated glassy carbon, the first activation energy is close to $E_1 \approx 23$ K, whereas amorphous $SiO_2$ exhibits almost four times the value. Consequently, the low-temperature plateau of $\kappa(T)$ in GO-based materials and rGC is shifted toward lower temperatures compared with $SiO_2$. These observations indicate that structural modifications primarily influence the relative strength of the coherence channels rather than their characteristic energy scales.

The multichannel analysis further reveals that not all coherence channels contribute equally to every material. The GO_300ºC_10t_S5 sample requires three activated channels to reproduce the measured thermal conductivity over the entire temperature interval, with characteristic energies satisfying approximately $E_2 \approx 3.8\ E_1$ and $E_3$

$\approx 23\ E_1$. By contrast, the GO_300ºC_S3, GO–HyFn_S1, and GO–HyFn_S2 samples exhibit no statistically significant second coherence channel, whereas the reticulated glassy carbon sample does not require the third channel. These observations indicate that the relative importance of the different coherence channels depends on the structural state of the material, even though their characteristic energy scales remain remarkably similar.

To examine whether this behavior extends beyond graphene oxide, Eq. (3) was also applied to expanded graphite and EG–MWCNT composites. Figure 5 presents the thermal conductivity of five EG and EG–MWCNT composites samples prepared by different oxidation routes and containing different densities and pore structures. Despite their distinct preparation histories, all samples exhibit the same qualitative temperature dependence: a monotonic increase in thermal conductivity above the inaccessible low-temperature plateau, followed by gradual saturation approaching room temperature. The incorporation of multiwalled carbon nanotubes modifies the absolute magnitude of the thermal conductivity but does not alter its overall temperature dependence.

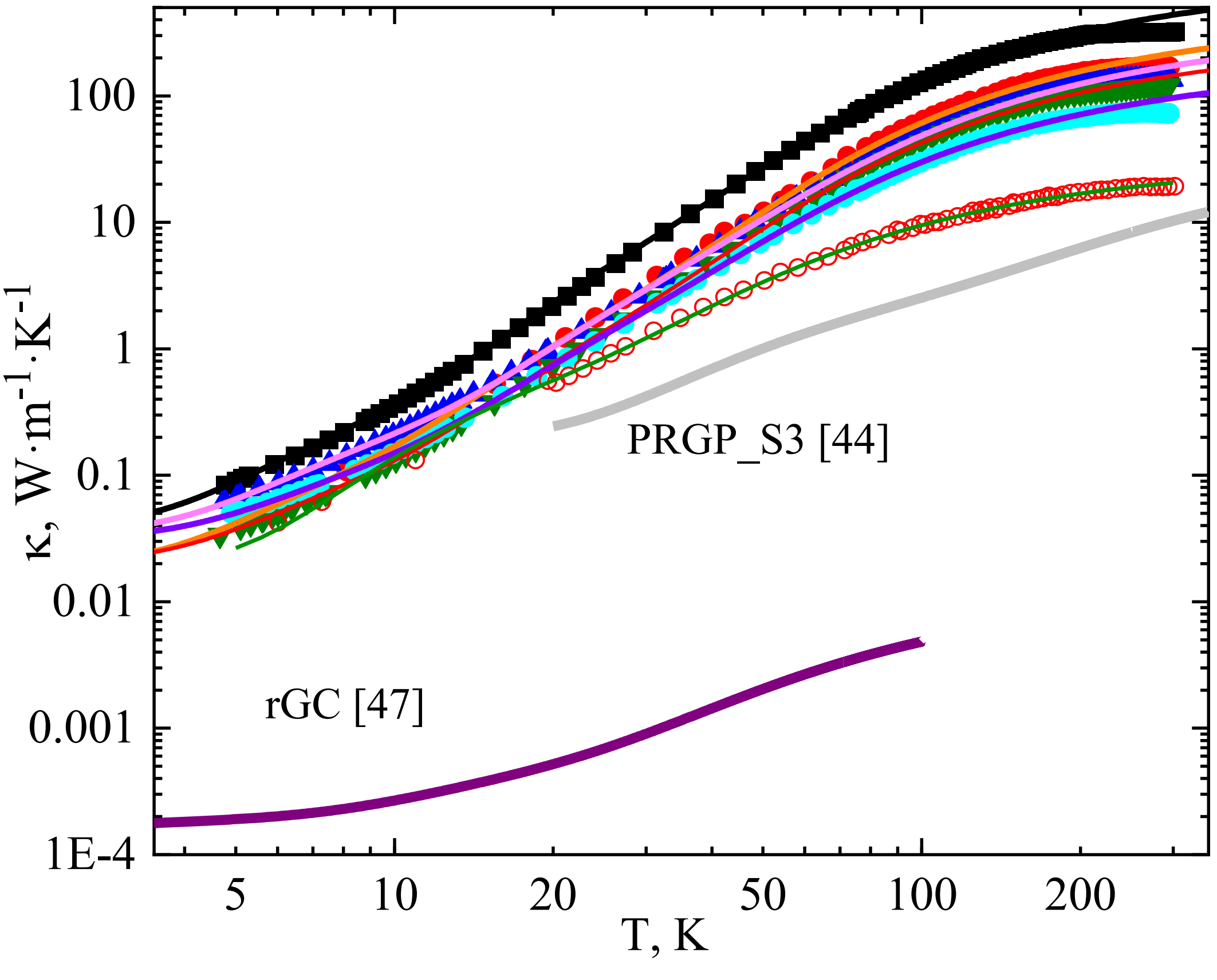

**Fig.5.** Temperature dependence of the thermal conductivity of expanded graphite (EG) and EG–MWCNT composites containing 3 wt.% multiwalled carbon nanotubes with different

densities ($\rho$) and sample thicknesses. Symbols: (●) – #6-Elechem$^{II-III}$, (▼) – #7-Chem$^{I}$, (▲) – #8-Chem$^{I}$, (●) – #13-Chem$^{I}$-3%CNTs, (■) – #14-Chem$^{I}$-3%CNTs. Literature data for graphene ceramics [4]: (○) GC_1000°C_5GPa and partially reduced graphene paper (PRGP_S3_round1) (—)[44] and reticulated glassy carbon (rGC) (—) [47] are shown as lines. Solid curves represent fits according to Eq. (3); the corresponding fitting parameters are listed in Table 2.

The solid curves in Fig. 5 demonstrate that Eq. (3) provides an excellent description of the thermal conductivity of all expanded graphite samples throughout the investigated temperature range. As observed for graphene oxide, the fitting procedure consistently identifies three thermally activated coherence channels. Most notably, the extracted activation energies exhibit nearly identical scaling relations for all investigated samples, with $E_2 \approx 4E_1$ and $E_3 \approx 12E_1$. The persistence of these characteristic energy ratios across materials possessing substantially different densities, pore morphologies, and interfacial structures strongly suggests that the observed multichannel behavior reflects a general property of structurally heterogeneous carbon allotropes rather than a peculiarity of a particular material.

Taken together, the results presented in Figs. 4 and 5 provide direct experimental evidence that the coherence contribution to thermal transport in carbon allotropes cannot, in general, be represented by a single activated process. Instead, the experimental data consistently reveal several characteristic activation-energy scales whose amplitudes depend on the material microstructure, whereas their characteristic energies remain remarkably robust across a broad range of carbon architectures.

### 4.4. Universality across carbon allotropes

The applicability of the multichannel phenomenological description extends well beyond graphene oxide. Figure 5 examines its validity for structurally distinct carbon allotropes by presenting the thermal conductivity of expanded graphite (EG) samples prepared using different oxidation routes together with EG–MWCNT composites spanning a range of densities. Literature data for partially reduced graphene paper (PRGP_S3_round1) [44] and reticulated glassy carbon (rGC) [47] are included for comparison.

Despite their pronounced differences in density, pore architecture, graphitic ordering, and interfacial connectivity, all of these materials exhibit remarkably similar thermal-transport characteristics. Their thermal conductivities evolve smoothly with temperature and differ primarily in magnitude rather than in functional form. This observation suggests that the governing transport mechanism is largely independent of the particular carbon architecture and instead reflects a common feature of structurally heterogeneous carbon networks.

As shown by the solid curves in Fig. 5, Eq. (3) accurately reproduces the thermal conductivity of all expanded graphite and EG–MWCNT samples over the entire investigated temperature range. The fitting reveals the same hierarchy of characteristic activation energies identified for graphene oxide, with $E_2 \approx (4{\div}5)\, E_1$ and $E_3 \approx (12{\div}15)\, E_1$. The persistence of these characteristic energy scales in materials prepared by different synthesis routes and possessing substantially different microstructures provides strong evidence that the multichannel description captures a general feature of thermal transport in disordered carbon materials rather than a peculiarity of graphene oxide.

The absolute magnitude of the thermal conductivity varies considerably among the investigated materials. For example, the thermal conductivity of reticulated glassy carbon is approximately two orders of magnitude lower than that of expanded graphite in the low-temperature region, whereas partially reduced graphene paper exhibits values approximately one order of magnitude lower than those of expanded graphite at higher temperatures. Nevertheless, after appropriate scaling, all materials display nearly the same overall temperature evolution, indicating that the dominant heat-transport processes remain qualitatively unchanged despite substantial differences in microstructure.

The solid curves in Fig. 5 show the results obtained using the multichannel expression given by Eq. (3). Excellent agreement with the experimental data is achieved for all expanded graphite and EG–MWCNT samples over the entire investigated temperature interval. As observed for graphene oxide, the thermal conductivity cannot be described adequately by a single activated coherence contribution. Instead, three coherence-mediated channels are required to reproduce the experimental temperature dependence.

A particularly important result is the remarkable reproducibility of the characteristic activation energies. For all expanded graphite samples, the lowest activation energy remains nearly constant, while the higher-energy channels satisfy the approximate scaling relations $E_2 \approx (4 \div 5)\, E_1$ and $E_3 \approx (12 \div 15)\, E_1$.

The persistence of these characteristic energy scales despite significant variations in density, oxidation route, and nanotube incorporation strongly suggests that they reflect intrinsic features of coherence-mediated vibrational transport rather than sample-specific fitting parameters.

The universality of the proposed phenomenological description is further supported by the analysis of previously published thermal-conductivity data for a wide variety of carbon materials. In addition to graphene oxide, graphene ceramics, expanded graphite and its composites with multiwalled carbon nanotubes, and reticulated glassy carbon, Eq. (3) successfully describes amorphous carbon thin films deposited by plasma-assisted chemical vapor deposition (PACVD) and filtered cathodic arc deposition reported in Ref. [48] – c.f. additional data in Table 2. For these materials, two coherence channels with activation energies falling within the ranges associated with $E_2$ and $E_3$ are sufficient to reproduce the experimental thermal conductivity.

By contrast, the thermal conductivity of tetrahedral and diamond-like amorphous carbon films reported in Ref. [49] is adequately described using only a single activated coherence contribution, whose characteristic energy lies within the range of $E_2$. This observation indicates that not every carbon allotrope necessarily exhibits all three coherence channels with measurable amplitudes. Instead, the number of experimentally observable channels depends on the structural state of the material and on the investigated temperature interval. Importantly, however, whenever additional channels become significant, their characteristic activation energies consistently fall within the same limited energy ranges.

Taken together, these observations demonstrate that the multichannel description possesses a remarkable degree of universality. Carbon allotropes differing by nearly three orders of magnitude in thermal conductivity, ranging from highly porous reticulated glassy carbon to dense expanded graphite and graphene-based materials, can all be interpreted within the same phenomenological framework. The principal distinction between these materials lies not in the characteristic activation energies themselves but in the relative amplitudes of the individual coherence channels, which reflect differences in structural disorder, interfacial connectivity, and network topology.

**4.5. Scaling laws**

The phenomenological parameters obtained from the multichannel analysis reveal several simple scaling relations that provide further insight into thermal transport in disordered carbon materials.

Figure 6 summarizes the dependence of the plateau thermal conductivity, $\kappa_{\mathrm{pl}}$, and the pre-exponential factors, $\kappa_{0i}$, on the density of the expanded graphite samples. A striking observation is that the plateau thermal conductivity remains essentially independent of density over the investigated range. In contrast, the amplitudes of the coherence channels exhibit a pronounced and systematic density dependence.

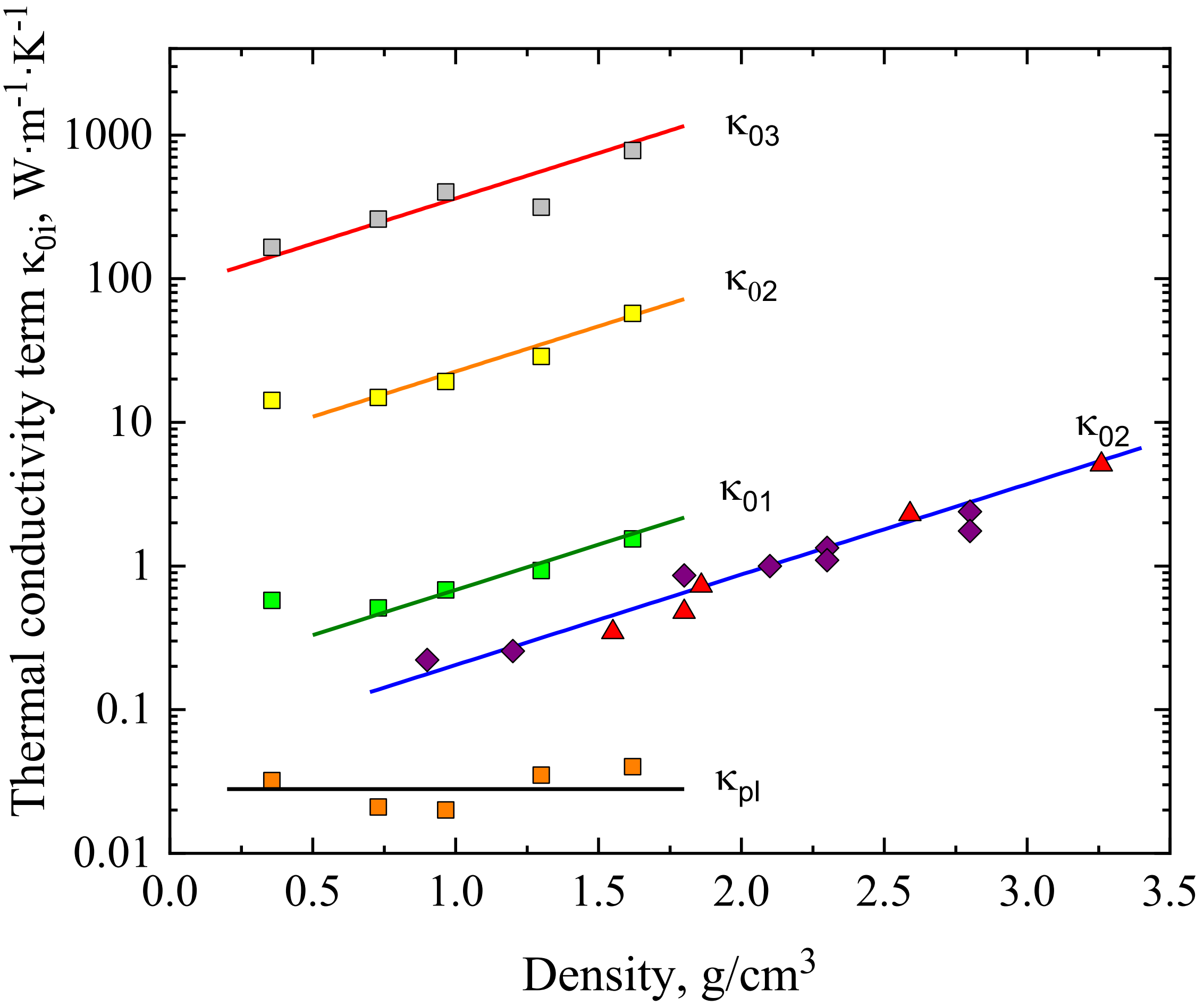

**Fig. 6.** Dependence of the pre-exponential factor $k_{0i}$ on the sample density $\rho$. Expanded graphite: symbols: (■) - $\kappa_{\mathrm{pl}}$, (■) – $\kappa_{01}$, (■) – $\kappa_{02}$, (■) – $\kappa_{03}$, (▲) – $\kappa_{02}$ for diamond-like carbon

films [49], (♦) - $\kappa_{02}$ for amorphous carbon thin films [48]. The colored solid lines represent fits to Eq. (4).

The pre-exponential factors increase exponentially with increasing density and are well described by

$$\kappa_{0i} = A_i \cdot \exp(1.45 \cdot \rho), \quad (4)$$

where $\rho$ is the sample density expressed in g cm$^{-3}$, and $A_i$ is the pre-factor associated with the $i$-th coherence channel. For the expanded graphite samples investigated in the present work, $A_1 = 0.16$ W m$^{-1}$K$^{-1}$, $A_2 = 5.3$ W m$^{-1}$K$^{-1}$, $A_3 = 85$ W m$^{-1}$K$^{-1}$. The three datasets form nearly parallel straight lines in semilogarithmic coordinates, indicating that increasing density amplifies all coherence channels by essentially the same factor while preserving their relative hierarchy.

To examine whether this behavior is specific to expanded graphite, the same analysis was performed using published thermal-conductivity data for diamond-like carbon films [49] and amorphous carbon thin films [48]. Although these materials span a density range approximately twice as large as that of the present samples, they exhibit essentially the same exponential dependence, yielding $A_2 = 0.048$ W m$^{-1}$K$^{-1}$ while retaining nearly the same exponential slope. This remarkable agreement suggests that density primarily controls the strength of coherence-mediated heat transport through a universal amplification mechanism that is largely independent of the particular carbon architecture.

The behavior of the characteristic activation energies is markedly different. Figure 7a shows the dependence of $E_i$ on sample density. For the expanded graphite samples, the activation energies form three well-defined groups corresponding to the three coherence channels, with no statistically significant dependence on density over the investigated range. The colored horizontal lines in Fig. 7a represent the corresponding average values.

A weak density dependence becomes apparent only when the broader dataset for diamond-like carbon and amorphous carbon films is included. In this case, the characteristic energy associated with the second coherence channel is approximately described by $E_2 = 38 \cdot \exp(0.35 \cdot \rho)$ as indicated by the dashed line in Fig. 7a. Even in this broader dataset, however, the variation of the activation energy remains substantially weaker than that of the corresponding pre-exponential factors. These observations indicate that structural densification primarily modifies the efficiency of the coherence channels rather than their intrinsic energy scales.

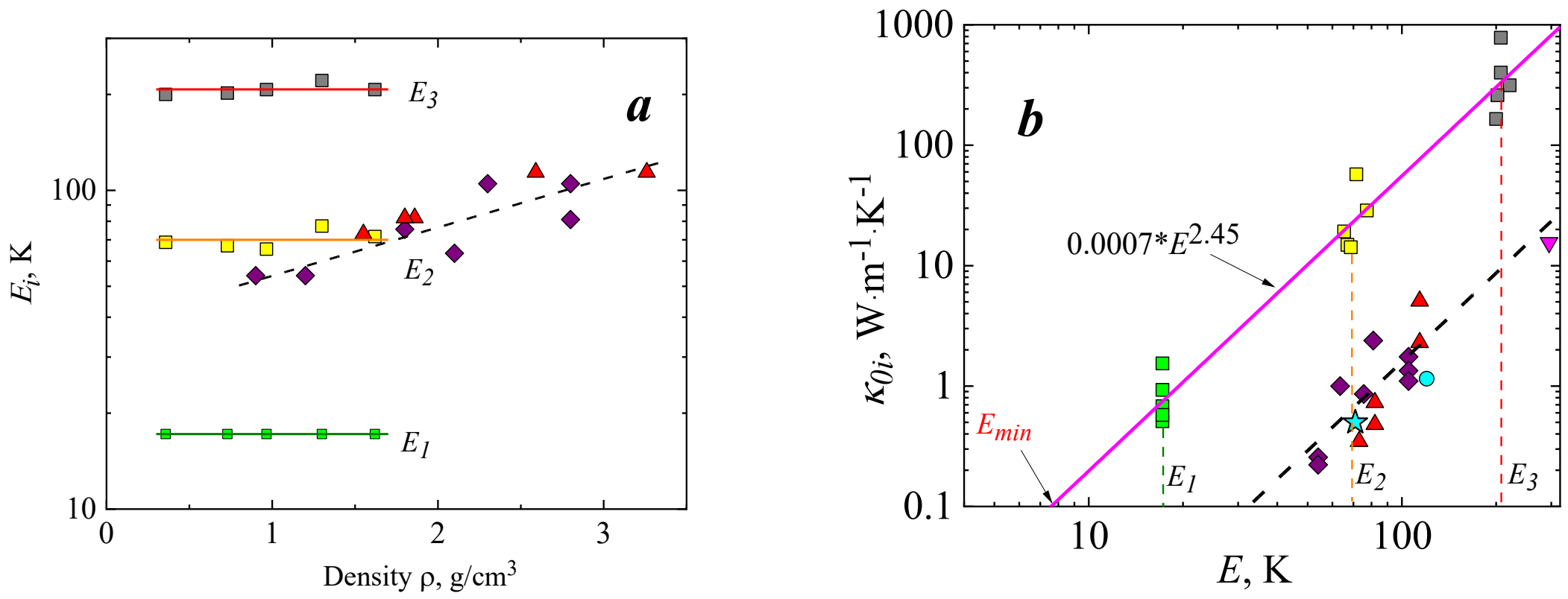

**Fig. 7. a)** Dependence of the characteristic energy $E_i$ on the density ($\rho$) of expanded graphite samples and diamond-like carbon (DLC) films [49] and amorphous carbon thin films [48]. The symbols are the same as in Fig. 6. **b)** Dependence of the pre-exponential factors $\kappa_{0i}$ on the characteristic energy $E_i$. The $\kappa_{02}(E_2)$ dependence for expanded graphite is compared with those for diamond-like carbon films [49], amorphous carbon thin films [48], (●) - amorphous carbon [1], (★) - partly-reduced graphene oxide film PRGO400 [45], (▼) - $\kappa_{03}(E_3)$ for glassy carbon GC [3]. The solid lines represent the fits described in the text.

An even more remarkable scaling relation emerges when the amplitudes of the coherence channels are plotted directly against their corresponding activation energies. Figure 7b presents the pre-exponential factors obtained for the expanded graphite samples together with literature values for amorphous carbon [1], partially reduced graphene oxide [45], glassy carbon [3], diamond-like carbon films [49], and amorphous carbon thin films [48].

For the expanded graphite samples, the data are accurately described by the power-law relation $\kappa_{0i} = 0.0007 \cdot E_i^{2.45}$ shown by the solid line in Fig. 7b. Extrapolation of this dependence suggests a threshold energy of approximately $E_{min} \approx 8$ K below which coherence-mediated transport becomes negligible.

Remarkably, the literature data follow the same scaling exponent, $\kappa_{02} = 0.00002 \cdot E_2^{2.45}$ differing primarily in the numerical prefactor. The persistence of the exponent over materials possessing widely different microstructures, densities, and absolute thermal conductivities strongly suggests that this power law reflects an intrinsic property of coherence-mediated thermal transport rather than a material-specific correlation.

Similar correlations between the activation energy and the pre-exponential factor have previously been reported for other classes of disordered solids. In graphene-filled epoxy composites [40], the coherence contribution obeys the Meyer–Neldel relation, $\kappa_0 = 0.011\exp(E/T_{\rm MN})$ with a Meyer–Neldel temperature $T_{\rm MN}$ = 13 K. Likewise, coherence-mediated thermal transport in molecular crystals and quasicrystals [50] exhibits a linear dependence of the pre-exponential factor on the activation energy, $\kappa_0(E) = \beta(E/E_{\rm min} - 1)$ where $\beta = 0.98 \pm 0.05$ W m$^{-1}$K$^{-1}$ and $E_{\rm min} = 165 \pm 10$ K for molecular compounds, and $\beta = 1.4 \pm 0.1$ W m$^{-1}$K$^{-1}$ and $E_{\rm min} = 46.7 \pm 3$ K for group of quasicrystals, where $E_{\rm min}$ is the threshold energy.

Although the mathematical forms of these correlations differ, they all indicate the existence of universal relationships linking the characteristic energy scale of coherence-mediated excitations with their contribution to thermal transport. The present results extend this concept to structurally heterogeneous carbon allotropes and demonstrate that both the density dependence of the amplitudes and the power-law relation between $\kappa_{0i}$ and $E_i$ are robust features of multichannel thermal transport.

**Table 2.** Fitting parameters describing the temperature dependence of the thermal conductivity of the investigated samples according to Eq. (3). Here, $\kappa_{\mathrm{pl}}$ denotes the plateau thermal conductivity, while $\kappa_{0i}$ and $E_i$ are the pre-exponential factors and characteristic energies, respectively, describing the thermal conductivity over the entire temperature range. The densities ($\rho$) of amorphous carbon thin films [48], diamond-like carbon (DLC) films [49], glassy carbon [3], partially reduced graphene paper, and graphene oxide paper [44] are included for comparison.

| Sample | $\kappa_{\mathrm{pl}}$ | $\kappa_{01}$ | $E_1$ | $\kappa_{02}$ | $E_2$ | $\kappa_{03}$ | $E_3$ | $\rho$ |
|---|---|---|---|---|---|---|---|---|
| | $W{\cdot}m^{-1}K^{-1}$ | $W{\cdot}m^{-1}K^{-1}$ | K | $W{\cdot}m^{-1}K^{-1}$ | K | $W{\cdot}m^{-1}K^{-1}$ | K | $g/cm^3$ |
| $SiO_2$ [31,46] | 0.12 | 1.37 | 89 | - | - | - | - | - |
| GO_300ºC_6t_S5 | 0.0125 | 0.198 | 23.2 | 0.275 | 89 | 2.91 | 538 | - |
| GO_300ºC_8t_S5 | 0.018 | 0.29 | 23.2 | 1.63 | 340 | - | - | - |
| GO_300ºC_10t_S5 | 0.0125 | 0.198 | 23.2 | 0.275 | 89 | 2.91 | 538 | - |
| GO_300ºC_S3 | 0.0159 | 0.22 | 23 | 0.02< | 90 | 2.8 | 490 | - |
| GO_500°C_10t | 0.018 | 0.29 | 23.2 | 0.808 | 106 | 3.25 | 605 | - |
| GO_700°C_10t | 0.0165 | 0.23 | 22.5 | 0.59 | 106 | 2.25 | 500 | - |
| GO-HyFn_S1 | 0.013 | 0.20 | 23.2 | 0.02< | 90 | 2.46 | 490 | - |
| GO-HyFn_S2 | 0.0136 | 0.22 | 25 | 0.02< | 90 | 2.8 | 490 | - |
| Graphene ceramics [4] | | | | | | | | |
| GC_1000°C_4GPa | 0.2 | 0.78 | 28.8 | 3.89 | 129 | 6.29 | 400 | - |
| GC_1000°C_5GPa | 0.02 | 2.10 | 28.8 | 28.7 | 129 | - | - | - |
| Expanded graphite and EG–MWCNT composites | | | | | | | | |
| #6-Elechem[II-III] | 0.032 | 0.575 | 17.2 | 14.2 | 68.7 | 165 | 200 | 0.357 |
| #7-Chem[I] | 0.021 | 0.51 | 17.2 | 14.9 | 67.0 | 260 | 202 | 0.73 |
| #8-Chem[I] | 0.035 | 0.93 | 17.2 | 28.7 | 77.3 | 314 | 221 | 1.3 |
| #13-Chem[I]-3%CNTs | 0.02 | 0.68 | 17.2 | 19.2 | 65.4 | 401 | 207 | 0.966 |
| #14-Chem[I]-3%CNTs | 0.04 | 1.54 | 17.2 | 57.26 | 71.6 | 780 | 207 | 1.62 |
| Amorphous carbon [1] | | | | | | | | |
| AC | - | 0.39 | 19 | 1.15 | 120 | 2.69 | 600 | - |
| *a*-C films, filtered-arc deposition [48] | | | | | | | | |
| H | - | - | - | 2.39 | 81 | 2.87 | 600 | 2.8 |
| I | - | - | - | 1.75 | 105 | 1.73 | 600 | 2.8 |
| L | - | - | - | 1.34 | 105 | 1.73 | 660 | 2.3 |
| M | - | - | - | 1.1 | 105 | - | - | 2.3 |
| *a*-C films deposited by plasma-assisted CVD [48] | | | | | | | | |
| E | - | - | - | 1 | 63.5 | 1.6 | 490 | 2.1 |
| B | - | - | - | 0.86 | 75.5 | 1.25 | 510 | 1.8 |
| F | - | - | - | 0.256 | 54 | 0.81 | 510 | 1.2 |
| C | - | - | - | 0.222 | 54 | 0.126 | 510 | 0.9 |
| The diamond-like carbon films [49] | | | | | | | | |
| tetrahedral amorphous carbon Ta_C Sbend | - | - | - | 5.09 | 114 | - | - | 3.26 |
| Ta_C HCA | - | - | - | 2.3 | 114 | - | - | 2.59 |
| DLCH (30%) | - | - | - | 0.73 | 82 | - | - | 1.86 |
| GLCH (18%) | - | - | - | 0.48 | 82 | - | - | 1.8 |
| PLCH (36%) | - | - | - | 0.347 | 73 | - | - | 1.55 |
| reticulated glassy carbon rGC [47] | 0.000198 | 0.0011 | 27 | 0.012 | 115 | - | - | - |
| glassy carbon GC [3] | - | - | - | - | - | 15.5 | 295 | 1.47 |
| Partly reduced graphene paper | - | 0.2 | - | 5.6 | 97 | 32 | 500 | 1.466 |
| PRGP_S3_round1 [44] | | | | | | | | |
| Graphene oxide paper GOP [44] | - | 0.7 | - | 3.6 | 290 | - | - | 1.357 |
| Partly-reduced graphene oxide film PRGO400 [45] | | | | 0.5 | 71 | 2.09 | 330 | |

### 4.6. Physical interpretation

The phenomenological analysis presented above demonstrates that the thermal conductivity of structurally heterogeneous carbon materials cannot generally be described by a single coherence contribution. Instead, the experimental data consistently reveal two or three characteristic activation-energy scales that govern the evolution of thermal conductivity above the low-temperature plateau. An important question therefore concerns the physical origin of these multiple coherence channels.

The present results suggest that multichannel behavior is a direct consequence of the hierarchical structural organization of porous carbon materials. Unlike conventional amorphous solids, graphene oxide, expanded graphite, reticulated glassy carbon, and related carbon allotropes possess a complex architecture extending over several length scales. Their structure consists of graphitic sheets, curved graphene fragments, pores of different sizes, folded lamellae, interparticle junctions, and weak mechanical contacts. Consequently, structural disorder is not characterized by a single length scale but rather by a broad hierarchy of geometrical and topological heterogeneities.

High-porosity carbon materials are particularly illustrative examples of such hierarchical systems. Most of their volume consists of empty space, while the solid phase forms a sparse framework of thin graphene sheets, graphitic flakes, or carbon nanotubes connected through relatively weak interparticle contacts. Heat therefore propagates not only within individual graphitic domains but also across a complex network of interfaces possessing widely different coupling strengths.

In these materials, mechanical densification does considerably more than simply reduce the pore volume. Compression continuously reorganizes the network of weakly connected graphene elements, increases the number and quality of interlayer contacts, modifies the vibrational freedom of individual structural units, and changes the connectivity of the entire carbon framework. Such geometrical reorganization alters the spectrum of low-energy collective excitations without fundamentally changing their characteristic energy scales. This interpretation is fully consistent with the experimental observation that the amplitudes of the coherence channels depend strongly on density, whereas the corresponding activation energies remain comparatively insensitive to structural modification.

From this viewpoint, the contributions $\kappa_{C1}$, $\kappa_{C2}$, and $\kappa_{C3}$ should not be regarded as independent microscopic heat-transfer mechanisms. Rather, they represent effective coherence-mediated transport channels associated with different groups of vibrational excitations emerging from the hierarchical topology of the carbon network. Each channel reflects a characteristic energy scale over which coherent vibrational coupling becomes effective in different parts of the heterogeneous structure.

This interpretation naturally explains why different carbon allotropes require different numbers of observable coherence channels. In materials possessing relatively simple structural organization, one or two channels may dominate the experimentally accessible temperature range. In contrast, highly heterogeneous graphene-derived materials contain a broader distribution of coupling strengths and characteristic length scales, allowing additional coherence channels to become experimentally distinguishable. The absence of a particular channel in some materials therefore should not be interpreted as evidence that the corresponding excitations are absent, but rather that their contribution to the measured thermal conductivity is too small to be resolved within the investigated temperature interval.

The present phenomenological picture is also consistent with recent theoretical developments in the description of thermal transport in disordered solids. The generalized Wigner transport formalism attributes heat transport to both particle-like phonon propagation and wave-like coherence between vibrational eigenstates. Likewise, recent atomistic simulations have shown that the thermal conductivity of disordered network solids is governed by the topology of the atomic-bond network and can be quantitatively related to bond-network entropy, which characterizes structural heterogeneity. Within this framework, increasing disorder modifies the distribution and coupling of vibrational states rather than simply increasing phonon scattering. The experimentally observed hierarchy of coherence channels may therefore represent the macroscopic manifestation of coherence-mediated transport within structurally heterogeneous vibrational networks.

Although the present analysis does not establish a unique microscopic origin for the three characteristic activation energies, it demonstrates that they constitute robust phenomenological energy scales recurring across chemically and structurally different carbon allotropes. Their remarkable reproducibility, together with the universal scaling relations established in the preceding section, strongly suggests that multichannel coherence transport is an intrinsic property of hierarchical carbon architecture rather than a consequence of a particular material or preparation method.

The emerging physical picture is therefore one in which thermal transport in porous carbon materials is governed by the interplay between geometry, structural disorder, and vibrational coherence across multiple length scales. Small structural modifications, such as mechanical compaction or changes in interlayer connectivity, can substantially alter the efficiency of the coherence channels while leaving their characteristic energy scales nearly unchanged. In this sense, thermal conductivity becomes an exceptionally sensitive probe of the hierarchical organization of disordered carbon networks, revealing how topology, dimensionality, and coherence collectively determine heat transport in these complex materials.

## 5. Conclusions

The thermal conductivity of a broad range of carbon allotropes, including graphene oxide, thermally reduced graphene oxide, GO–hydrated fullerene composites, graphene ceramics, expanded graphite, and expanded graphite–MWCNT composites, has been investigated in the temperature range 4.5–300 K. Although all materials exhibit glass-like thermal-conductivity behavior, their temperature dependences cannot be adequately described by the conventional two-channel model consisting of a Peierls contribution and a single thermally activated coherence term.

To describe the experimental data over the entire temperature range, a generalized multichannel phenomenological model has been proposed, in which the coherence contribution is represented by up to three Arrhenius-type terms characterized by distinct activation energies. The characteristic energies exhibit a remarkable hierarchy, approximately satisfying ($E_2 \approx (4 \div 5)\, E_1$ and $E_3 \approx (12 \div 15)\, E_1$), while their amplitudes depend strongly on the structural state of the material.

Analysis of both the present measurements and published data for a variety of carbon materials demonstrates that this multichannel description is applicable across structurally diverse carbon allotropes. The observed universality, together with the scaling relations between the phenomenological parameters, suggests that coherence-mediated heat transport is governed primarily by the hierarchical organization of disordered carbon networks rather than by the specific carbon structure.

These results provide experimental evidence that thermal transport in heterogeneous carbon materials proceeds through multiple coherence-mediated channels rather than a single activated process. The proposed phenomenological framework offers a simple and robust description of thermal transport in disordered carbon systems and establishes a basis for future theoretical investigations of coherence-mediated heat conduction in hierarchically disordered solids.

**Acknowledgments**

The work of A.I.K., V.V.S, Yu.V.H., S.V.C., & N.A.V. was supported by the National Research Foundation of Ukraine (Project №. 2023.03/0012). The work of A.J. and D.S. was supported by National Science Centre Poland (grant no. 2022/45/B/ST3/02326)

**Conflict of interest**

The authors declare no conflict of interest.

**Data accessibility statement**

The data that support the findings of this study are available upon reasonable request from the authors.